\documentclass[journal,comsoc]{IEEEtran}

\usepackage{cite}
\usepackage{amsmath,amssymb,amsfonts}
\usepackage{algorithm}
\usepackage{algpseudocode}
\usepackage{graphicx}
\usepackage{textcomp}
\usepackage{xcolor}
\usepackage{verbatim}
\usepackage{color}
\usepackage{colortbl}

\def\BibTeX{{\rm B\kern-.05em{\sc i\kern-.025em b}\kern-.08em
    T\kern-.1667em\lower.7ex\hbox{E}\kern-.125emX}}
\begin{document}
\title{Self-Interference Cancellation for Full-Duplex Massive MIMO OFDM with Single RF Chain}
\author{Meng He,~\IEEEmembership{Student Member,~IEEE,} and Chuan Huang,~\IEEEmembership{Member,~IEEE}}
\maketitle
\begin{abstract}
In this paper, the digital self-intereference (SI) canellation in a single radio frequency (RF) chain massive multi-input multi-output (MIMO) full-duplex (FD) orthogonal frequency division multiplexing (OFDM) system with phase noise is studied. To compsenate the phase noise, which introduces SI channel estimation error and thus degrades the SI cancellation performance, a weighted linear SI channel estimator is derived to minimize the residual SI power in each OFDM symbol. The digital SI cancellation ability of the proposed method, which is defined as the ratio of the SI power before and after the SI cancellation, is anlyzed. Simulation results show that the proposed optimal linear SI channel estimator significantly outperforms the conventional least square (LS) etimator in terms of the SI cancellation ability for the cases with strong SI and low oscillator quality. 
\end{abstract}
\begin{IEEEkeywords}
Full-duplex,  massive MIMO, self-interference cancellation, channel estimation
\end{IEEEkeywords}
\footnote{The authors are with the Future Network of Intelligence Institute, The Chinese University of Hong Kong (Shenzhen), China (Emails: victorhe96@outlook.com and huangch@uestc.edu.cn)}
\section{Introudction}A
Full duplex (FD) technology has the potential to double the spectrical efficiency of a communicaotn system by allowing the transmitter and the receiver to simultaneously work at the same frequency band \cite{6319352}. However, the main challenge of a practical FD system is to efficiently cancel the strong self-intereference (SI) from its local transmitter to the local receiver \cite{7815419}. Generally, the SI cancellation in the FD communication systems is implemented in the following three parts: the antenna domain \cite{6702851}, analog domain \cite{6523998}, and digital domain \cite{6542771}. 

For the digital SI cancellation schemes, phase noise introduced by the imperfect oscillators at the transmitter and the receiver, is concluded to be the bottlelock of the digital SI cancellation performance \cite{7815419}, since it causes SI channel estimation error and
residual SI \cite{6523998,6542771,8403642,6937196}. In \cite{6212402}, the channel and phase noise are jointly estimated for the multi-input multi-output (MIMO) systems, where the channel is with a single tap. However, in practical FD systems, multipath SI channel should be considered, and minimizing the SI channel estimation error may not achieve the best SI cancellation performance in terms of the SI cancellation ability \cite{8403642,123456,234567}.

In this paper, we consider the digital SI cancellation in a single RF chain FD massive MIMO OFDM systems with phase noise at both the transmitter and the receiver. A linear wegithed SI channel estimator is derived to minmize the power of the residual SI. The closed-form expression for the power of the residual SI is formulated, and the optimal linear SI channel estimator is computed by solving the residual SI power minimization problem. Finally, the SI cancellation ability is defined and calculated for the proposed optimal linear SI channel estimator.

Notations: Bold face small and big letters, e.g. $\mathbf{x}$ and $\mathbf{X}$, denote vectors and matrices, respectively. $\Re\{\mathbf{X}\}$, $\Im\{\mathbf{X}\}$, $(\mathbf{X})^T$, $(\mathbf{X})^H$, $\mathrm{tr}\{\mathbf{X}\}$, and $[\mathbf{X}]_{i,j}$ are the real part, imaginary part, transpose, conjugate transpose, trace, and the $(i,j)$-th entry of the matrix $\mathbf{X}$. $\Vert\cdot\Vert^2$ is the L-2 norm, and $\mathbb{E}[\cdot]$ represents the expectation.
\section{System Model}
In this section, the OFDM based communications in a full-duplex(FD) Massive MIMO system is considered, where the base station (BS) has $N_s$ transmit antennas and $N_r$ receive antennas, with a single radio frequency (RF) chain. At the local transmitter of the BS, the transmit digital symbols $\{X[k]\}_{k=0}^{N_c-1}$ are first transformed into the time domain signal $x(t)$ by a standard $N_c$ point OFDM modulation with $N_c$ being the total number of subcarriers in one OFDM symbol. Then, at the local transmit oscillator, $x(t)$ is upconverted to construct the local RF transmit signal. The RF signal at the $s$-th  transmit antenna is given as \cite{7815419}
\begin{equation}
\label{transmit signal 1}
\tilde{x}_s(t)=x(t)e^{j(2\pi f_ct +\theta_s^T(t))},
\end{equation}
where $f_c$ is the carrier frequency, and $\theta_s^T(t)$ represents the phase noise introduced by the imperfect oscillator at the $s$-th transmit antenna. Especially, the phase noise in all oscillators at both the transmitter and the receiver are modeled as Wiener process with variance $\sigma_{\theta}^2=4\pi \Delta f$, where the relative bandwidth $\Delta f$ of the phase noise is a widely adopted oscillator quality evaluation factor\cite{6937196}.

While the RF singal defined in \eqref{transmit signal 1} is transmitted from the BS to the remote users, it is also received by the local receiver at the BS via a multipath channel. Thus, the recevied RF signal at the BS consists of both the signal-of-interest (SoI) from the remote users and the self-interference from local transmitter, i.e., the received signal at the $r$-th antenna is given as \cite{8403642}
\begin{equation}
\label{received RF signal at r-th antenna}
\tilde{y}_r(t)=\sum_{s=1}^{N_s}\sum_{l=0}^{L-1}h_{s,r}(l)\tilde{x}_s(t-\tau_l)+y_r^U(t)+n_r(t),
\end{equation}
where the first term in the right hand side of \eqref{received RF signal at r-th antenna} represents the SI signal, $\{h_{s,r}(l)\}_{l=0}^{L-1}$ is the multipath SI channel coefficients of the link between the $s$-th transmit antenna and $r$-th receive antenna of the BS, $L$ is the total number of multipath channel taps, $\tau_l$ dentoes the transmission delay, $y_r^U(t)$ is the received SoI from the uplink transmission, and $n_r(t)$ is the circular symmetric complex Gaussian (CSCG) noise with zero mean and variance $\sigma_n^2$.

At the local receiver oscillator of $r$-th antenna, $\tilde{y}_r(t)$ is first downconverted to the baseband, where this baseband signal is given as \cite{7815419}
\begin{equation}
y_r(t)=\tilde{y}_r(t)e^{-j2\pi(f_ct-\theta_r^R(t))},
\end{equation}
where $\theta_r^R(t)$ is the phase noise at the $r$-th receive antenna. Then, with OFDM demodulation, the time domain signal $y_r(t)$ is restored to the digital frequency domain symbols $\{Y_r[k]\}_{k=0}^{N_c-1}$ \cite{6937196}, i.e.,
\begin{equation}
\label{digital frequency domain signal at r-th antenna}
Y_r[k]
=\sum_{s=1}^{N_s}\sum_{i=0}^{N_c-1}\delta_{i-k}^{s,r}H_{s,r}[i]X[i]+Y_{r}^{U}[k]+N_r[k],
\end{equation}
where $\{X[k]\}_{k=0}^{N_c-1}$ are the transmit digital symbols, $\{H_{s, r}[k]\}_{k=0}^{N_c-1}$ are the frequency domain SI channel coefficients, $Y_r^U[k]$ is the digital frequency domain SoI symbol, $N_r[k]$ is the frequency domain CSCG noise, and $\delta_{i-k}^{s,r}$ is the frequency domain SI phase noise consisting of the impacts of both the transmitter phase noise $\theta_s^T(t)$ and receiver phase noise $\theta_r^R(t)$ \cite{7815419}.

\section{Optimal Linear SI Channel Estimator}
In this section, a weighted linear SI channel estimator is derived to minmize the power of the residual SI. The closed-form expression for the power of the residual SI is formulated, and the linear channel estimator is optimized by solving the residual SI power minmization problem. Finally, the digital SI cancellation ability of the proposed method is defined and calculated.
\subsection{SI Channel Estimation and SI Cancellation}
By \eqref{digital frequency domain signal at r-th antenna}, the matrix form of the received frequency domain signal for one OFDM symbol is expressed as 
\begin{equation}
\label{matrix form of the received signal 1}
    \mathbf{y}_r=\mathbf{XF}\sum_{s=1}^{N_S}\delta_{0}^{s,r}
    \mathbf{h}_{s,r}+\mathbf{m}_r+\mathbf{y}_r^U+\mathbf{n}_r,
\end{equation}
where  
$\mathbf{X}=\mathrm{diag}\{X[0],X[1],\cdots,X[N_c-1]\}$ is the SI symbol matrix, $\mathbf{h}_{s,r}=[h_{s,r}(0),$ $h_{s,r}(1), \cdots, h_{s,r}(l-1)]^T$ is the SI channel impulse vector, $\mathbf{y}_r^U=[Y_r^U[0], Y_r^U[1], \cdots, Y_r^U[N_c-1]]^T$ is the SoI vector, $\mathbf{n}_r=[N_r[0],N_r[1],\cdots,N_r[N_c-1]]^T$ is the additive CSCG noise vector, $\mathbf{m}_r$ is the inter-carrier interference (ICI) \cite{6937196} vector with its entries being 
$$[\mathbf{m}_r]_k=\sum_{i=0,i\neq k}^{N_c-1}X[i]\sum_{s=1}^{N_s}\delta_{i-k}^{s,r}H_{s,r}[i], \,k=0,1,\cdots,N_c-1,$$
and $\mathbf{F}$ is a $N_c\times L$ discrete fourier transform matrix \cite{7815419} with its entries being 
$$[\mathbf{F}]_{n,l}=e^{-j2\pi nl/N_c}.$$
Since the SI cancellation at each receive antenna is similar, the script $r$ in \eqref{matrix form of the received signal 1}, which denotes the index of the receive antenna, is omitted for simplicity in the sequel. 

Denote $\mathbf{h}_{\delta}=\sum_{s=1}^{N_s}\delta_{0}^s\mathbf{h}_{s}$ as the combined SI channel influence rotated with the common phase error (CPE) \cite{6937196}, a weighted linear SI channel estimator of $\mathbf{h}_{\delta}$ is proposed, i.e.
\begin{equation}
\label{SI channel estimator 1}
	\hat{\mathbf{h}}_{\delta}=\mathbf{Wy}=\mathbf{WXF}\hat{\mathbf{h}}_{\delta}+\mathbf{W}\mathbf{m}+\mathbf{W}\mathbf{y}^U+\mathbf{Wn},
\end{equation}
where $\mathbf{W}$ is a $L\times N_c$ weight matrix to minimize the residual SI power. Then, using the SI channel estimator $\hat{\mathbf{h}}_{\delta}$ in \eqref{SI channel estimator 1}, the SI signal is reconstructed as 
\begin{equation}
     \hat{\mathbf{y}}^I=\mathbf{XF}\hat{\mathbf{h}}_{\delta}=
     \mathbf{XFW}(\mathbf{XF}\mathbf{h}_{\delta}
     +\mathbf{m}+\mathbf{n}+\mathbf{y}^U).
\end{equation}
By substracting $\hat{\mathbf{y}}^I$ from the received signal $\mathbf{y}$ in \eqref{matrix form of the received signal 1},  the recovered SoI is given as 
\begin{equation}
    \hat{\mathbf{y}}^U=\mathbf{y}-\hat{\mathbf{y}}^I=\mathbf{y}^U+\mathbf{r},
\end{equation}
 where
\begin{equation}
\label{residual SI 1}
\mathbf{r}=(\mathbf{I}_{N_c}-\mathbf{XFW})(\mathbf{XF}\mathbf{h}_{\delta}+\mathbf{m}+\mathbf{n})-\mathbf{XFW}\mathbf{y}^U,
\end{equation} 
is the residual SI with $\mathbf{I}_{N_c}$ being an $N_c$ order identity matrix. For simplification, we define a new $N_c\times N_c$ weight matrix  $\mathbf{V}=\mathbf{XFW}$. Since the diagonal matrix $\mathbf{X}$ in \eqref{matrix form of the received signal 1} is full rank, we obtain $\mathbf{W}$ from $\mathbf{V}$ by 
\begin{equation}
\label{wegithed matrix 1}
   \mathbf{W}=\frac{1}{N_c}\mathbf{F}^H\mathbf{T}\mathbf{V},
\end{equation}
where $\mathbf{T}=(\mathbf{X}^H\mathbf{X})^{-1}\mathbf{X}^H$ is the inverse matrix of $\mathbf{X}$. We also define $\mathbf{y}^I=\mathbf{XF}\mathbf{h}_{\delta}+\mathbf{m}$ as the SI signal before the SI cancellation. Then, the residual SI in \eqref{residual SI 1} is rewritten as
\begin{equation}
\label{residual SI 2}
\mathbf{r}=(\mathbf{I}_{N_c}-\mathbf{V})(\mathbf{y}^I+\mathbf{n})-\mathbf{V}\mathbf{y}^U.	
\end{equation}

\subsection{Optimal SI Channel Estimator}
The wegiht matrix $\mathbf{V}$ is designed to minimize the residual SI power in each OFDM symbol. Using \eqref{residual SI 2}, the power of the residual SI with known SI symbol matrix $\mathbf{X}$ in one OFDM symbol is given as 
\begin{equation}
\label{power of the residual SI 1}
\begin{split}
E_{r|\mathbf{X}}
=&\mathbb{E}[\Vert(\mathbf{I}_{N_c}-\mathbf{V})(\mathbf{y}^I+\mathbf{n})-\mathbf{V}\mathbf{y}^U\Vert^2|\mathbf{X}].
\end{split}
\end{equation}
Since the SI signal $\mathbf{y}^I$, signal-of-interest $\mathbf{y}^U$, and CSGN noise $\mathbf{n}$ in \eqref{power of the residual SI 1} are independents of each other \cite{7815419}, \eqref{power of the residual SI 1} is simplified as
\begin{equation}
\label{power of the residual SI 2}
\begin{split}
	E_{r|\mathbf{X}}
	=&\mathbb{E}[\Vert(\mathbf{I}_{N_c}-\mathbf{V})\mathbf{y}^I\Vert^2|\mathbf{X}]
	+\mathbb{E}[\Vert(\mathbf{I}_{N_c}-\mathbf{V})\mathbf{n}\Vert^2|\mathbf{X}]\\
	&+\mathbb{E}[\Vert\mathbf{V}\mathbf{y}^U\Vert^2|\mathbf{X}].
\end{split}
\end{equation}
The first term of $E_{r|\mathbf{X}}$ in \eqref{power of the residual SI 2} is calculated as
\begin{equation}
\label{first term of 10}
\begin{split}
E_1=&\mathbb{E}[\Vert(\mathbf{I}_{N_c}-\mathbf{V})\mathbf{y}^I\Vert^2|\mathbf{X}]\\
=&\mathrm{tr}\{\mathbb{E}[\mathbf{y}^I(\mathbf{y}^I)^H|\mathbf{X}]\}-2\Re\{\mathrm{tr}\{\mathbf{V}\mathbb{E}[\mathbf{y}^I(\mathbf{y}^I)^H|\mathbf{X}]\}\}\\
&+\mathrm{tr}\{\mathbf{V}\mathbb{E}[\mathbf{y}^I(\mathbf{y}^I)^H|\mathbf{X}]\mathbf{V}^H\}.
\end{split}
\end{equation}
The second term in \eqref{power of the residual SI 2} is calculated as
\begin{equation}
\begin{split}
\label{second term of 10}
E_2=&\mathbb{E}[\Vert(\mathbf{I}_{N_c}-\mathbf{V})\mathbf{n}\Vert^2]\\
=&N_c\sigma_n^2
-2\sigma_n^2\Re\{\mathrm{tr}\{\mathbf{V}\}\}
+\sigma_n^2\mathrm{tr}\{\mathbf{V}\mathbf{V}^H\},
\end{split}
\end{equation}
The third term in \eqref{power of the residual SI 2} is calculated as 
\begin{equation}
\label{third term of 10}
E_3=\mathbb{E}[\Vert\mathbf{V}\mathbf{y}^U\Vert^2]=
\sigma_u^2\mathrm{tr}\{\mathbf{V}\mathbf{V}^H\},
\end{equation}
where $\sigma_u^2=\mathbb{E}[|Y^U[k]|^2]$ is the power of SoI.

Let  $\mathbf{A}=\mathbb{E}[\mathbf{y}^I(\mathbf{y}^I)^H|\mathbf{X}]$ denote the convariance matrix of the SI signal with known $\mathbf{X}$ in one OFDM symbol, the entries of the convariance matrix $\mathbf{A}$ is calculated as
\begin{equation}
\label{matrix A 1}
\begin{split}
[\mathbf{A}]_{m,n}=
&\mathbb{E}[[\mathbf{y}^I(\mathbf{y}^I)^H]_{m,n}|\mathbf{X}]\\
=&\sum_{s=1}^{N_s}\sum_{i=0}^{N_c-1}\sum_{j=0}^{N_c-1}\bigg(\mathbb{E}[\delta_{i-m}^{s}\delta_{j-n}^{s^*}]X[i]X[j]^*\\
&\cdot\sum_{l=0}^{L-1}\mathbb{E}[|h_s(l)|^2]e^{-j\frac{2\pi}{N_c} (i-j)l}\bigg),
\end{split}
\end{equation}
where the phase noise convariance $\mathbb{E}[\delta^s_{i-m}\delta^{s^*}_{j-n}]$ depends on the osillator quality and types \cite{7815419}.
Notice  $\mathbf{A}=\mathbb{E}[\mathbf{y}^I(\mathbf{y}^I)^H|\mathbf{X}]=\mathbb{E}[(\mathbf{y}^I(\mathbf{y}^I)^H)^H|\mathbf{X}]=\mathbf{A}^H$, and thus $\mathbf{A}$ is a hermitian matrix with $\Re\{\mathbf{A}\}=\Re\{\mathbf{A}^T\}$ and $ \Im\{\mathbf{A}\}=-\Im\{\mathbf{A}^T\}$. Using \eqref{first term of 10}-\eqref{third term of 10}, \eqref{power of the residual SI 2} is rewritten as 
\begin{equation}
\label{residual SI power 2}
\begin{split}
E_{r|\mathbf{X}}=N_c\sigma_n^2+\mathrm{tr}\{\mathbf{A}\}
+\mathrm{tr}\{\mathbf{V}\mathbf{C}\mathbf{V}^H\}
-2\Re\{\mathrm{tr}\{\mathbf{V}\mathbf{B}\}\},
\end{split}
\end{equation}
where $\mathbf{B}=\mathbf{A}+\sigma_n^2\mathbf{I_{N_c}}$, and $\mathbf{C}=\mathbf{A}+\sigma_n^2\mathbf{I_{N_c}}+\sigma_u^2\mathbf{I_{N_c}}$.

With the closed-form expression for the residual SI power $E_{r|\mathbf{X}}$ in \eqref{residual SI power 2}, the optimization of the weigthed linear SI channel estimator is equivlent to designing the weight matrix $\mathbf{V}$ to minimize $E_{r|\mathbf{X}}$, i.e.,
\begin{equation}
\label{minimization problem 2}
\mathop{\mathrm{min}}\limits_{\{\mathbf{V}\}}\quad
E_{r|\mathbf{X}}.
\end{equation}
Notice \eqref{residual SI power 2} can be rewritten as 
\begin{equation} 
\label{residual SI 3}
E_{r|\mathbf{X}}=N_c\sigma_n^2+\mathrm{tr}\{\mathbf{A}\}
+\sum_{k=0}^{N_c-1}(\mathbf{v}_k^T\mathbf{\Phi}\mathbf{v}_k-2\mathbf{b}_k^T\mathbf{v}_k),
\end{equation}
where $\mathbf{v}_k$ is a $2N_c\times 1$ real vector given by
\begin{equation}
\label{v 1}
\begin{split}
\mathbf{v}_k=[&\Re\{[\mathbf{V}]_{k,0}\},\Re\{[\mathbf{V}]_{k,1}\},
\cdots,\Re\{[\mathbf{V}]_{k,N_c-1}\},\\
&\Im\{[\mathbf{V}]_{k,0}\},\Im\{[\mathbf{V}]_{k,1}\}, \cdots, \Im\{[\mathbf{V}]_{k,N_c-1}\}]^T,
\end{split}
\end{equation}
$\mathbf{\Phi}$ is a real block martix, i.e.
\begin{equation}
\label{Phi 1}
\begin{split}
\mathbf{\Phi}=\begin{bmatrix}
\Re\{\mathbf{C}\}&\Im\{\mathbf{C}\}\\
-\Im\{\mathbf{C}\}&\Re\{\mathbf{C}\}\\
\end{bmatrix},
\end{split}
\end{equation}
and $\mathbf{b}_k$ is a real vector with its entries being
\begin{equation}
\label{u 1}
\begin{split}
\mathbf{b}_k=[&\Re\{[\mathbf{B}]_{0,k}\},\Re\{[\mathbf{B}]_{1,k}\},
\cdots,\Re\{[\mathbf{B}]_{N_c-1,k}\},\\
&-\Im\{[\mathbf{B}]_{0,k}\},-\Im\{[\mathbf{B}]_{1,k}\}, \cdots, -\Im\{[\mathbf{B}]_{N_c-1,k}\}]^T.
\end{split}
\end{equation}
For \eqref{residual SI 3}, since both the first term $N_c\sigma_n^2$ and the second term $\mathrm{tr}\{\mathbf{A}\}$ are independent of $\mathbf{V}$, the minimization of $E_{r|\mathbf{X}}$ is identical to the minimization of the third term of $E_{r|\mathbf{X}}$. Thus, the minimizztion problem in \eqref{minimization problem 2} is equivalent to $N_c$ independent quadratic programming problems, i.e.,
\begin{equation}
\label{minimization problem 1}
\mathop{\mathrm{min}}\limits_{\{\mathbf{v}_k\}}\quad
\mathbf{v}_k^T\mathbf{\Phi}\mathbf{v}_k-2\mathbf{b}_k^T\mathbf{v}_k,
\end{equation}
and the optimization solution $\mathbf{v}_k^{*}$ and optimal value $f_k^*$ of each problem are given as 
\begin{equation}
\label{optimal solution and optimal value 1}
\begin{split}
\mathbf{v}_k^*=&\mathbf{\Phi}^{-1}\mathbf{b}_k,\\
f_k^*=&-\mathbf{b}_k^T\mathbf{\Phi}^{-1}\mathbf{b}_k,
\end{split}
\end{equation}
where the inverse matrix of $\mathbf{\Phi}$ is given as
\begin{equation}
\mathbf{\Phi}^{-1}=\begin{bmatrix}
\Re\{\mathbf{C}^{-1}\}&\Im\{\mathbf{C}^{-1}\}\\
-\Im\{\mathbf{C}^{-1}\}&\Re\{\mathbf{C}^{-1}\}\\
\end{bmatrix}.
\end{equation}
\begin{algorithm}[htbp]
	\caption{Optimal Linear SI channel Estimator}
	\label{Wegihted SI channel Estimator Optimization} 
	\begin{algorithmic}[1]
		\State
		Calculate matrix $\mathbf{A}$, $\mathbf{B}$, and $\mathbf{C}$ by \eqref{matrix A 1}.	
		\State
		Calculate matrix $\mathbf{\Phi}$ and vector $\{\mathbf{b}_k\}_{k=0}^{N_c-1}$ according to \eqref{Phi 1} and \eqref{u 1}.
		\State
		Calculate $\{\mathbf{v}_k\}_{k=0}^{N_c-1}$ by \eqref{optimal solution and optimal value 1} and obtain the optimal weight matrix $\mathbf{V}^*$ by $\{\mathbf{v}_k\}_{k=0}^{N_c-1}$.
		\State 
		Using $\mathbf{V}^*$ to obtain the optimal SI channel estimator by \eqref{SI channel estimator 1} and \eqref{wegithed matrix 1}. 
	\end{algorithmic}
\end{algorithm}

In summary, the optimization of the linear SI channel estimator is expressed in Algorithm 1, where it finds the optimal weight matrix $\mathbf{V}$ to minimize the residual SI power in each OFDM symbol. 
\subsection{SI Cancellation Ability}
\label{SI Cancellation Ability}
In this subsection, the performance of the proposed optimal linear  SI channel estimator is measured in terms of the digital SI cancellation ability\cite{7815419}. First, we define the SI cancellation ability as
\begin{equation}
\label{SI cancellation ability}
G=\frac{E_I+N_c\sigma_n^2}{E[E_{r|\mathbf{X}}]},
\end{equation}
where $E_I=\mathbb{E}[\Vert\mathbf{y}^I\Vert^2]=N_c\sigma_x^2\sum_{s=1}^{N_s}\sum_{l=0}^{L-1}\mathbb{E}[|h_{s}(l)|^2]$ is the SI power in one OFDM symbol before the SI cnacellation with $\sigma_x^2=\mathbb{E}[|X[k]|^2]$ being the power of the transmit symbol, and $E_r=\mathbb{E}[E_{r|\mathbf{X}}]$ is the residual SI power after the SI cancellation. 

Using \eqref{residual SI 3} and \eqref{optimal solution and optimal value 1}, the minimum residual SI power of the proposed linear SI channel estimator with knonw $\mathbf{X}$ in each OFDM symbol is given as 
\begin{equation}
E_{\mathrm{min}|\mathbf{X}}=N_c\sigma_n^2+\mathrm{tr}\{\mathbf{A}\}
+\sum_{k=0}^{N_c-1}f_k^*.
\end{equation}
With $E_{\mathrm{min}|\mathbf{X}}$, the cancellation ability of the proposed optimal linear SI channel estimator is computed as 
\begin{equation}
\label{maximun SI cancellation ability 1}
\begin{split}
G_{\mathrm{max}}=&\frac{E_I+N_c\sigma_n^2}{E[E_{\mathrm{min}|\mathbf{X}}]}\\
=&\frac{E_I+N_c\sigma_n^2}{E_I+N_c\sigma_n^2+\sum_{k=0}^{N_c-1}E[f_k^*]}.
\end{split}
\end{equation}
\section{Simulation and Numberical Results}
In this section, the digital SI cancellation ability of the proposed method and the conventional LS based method \cite{969514} are compared for interference-to-noise ratio (INR) defined as $\mathrm{INR}=E_I/N_c\sigma_n^2$, signal-to-noise ratio (SNR) defined as $\mathrm{SNR}=\sigma_u^2/\sigma_n^2$, and relative phase noise bandwidth $\Delta f$. Besides, the theoretical SI cancellation ability of the proposed method is calculated to verify the analytical results in section \ref{SI Cancellation Ability}.

At the transmitter, binary-phase-shift-keying (BPSK) modulation is adopted. Other parameters are set as: total number of transmit antennas $N_s=64$, OFDM subcarriers $N_c=128$, subcarrier spcing $f_{\mathrm{sub}}=$ 15 kHz, sample time $T_s=1/(f_{sub}\cdot N_c)=5\times10^{-7}$ s, and the OFDM cyclic prefix $N_p=16$. Both the phase noise at the transmitter oscillator and receiver oscillator are modeled as Wiener process with variance $\sigma_{\theta}^2=4\pi \Delta f$ \cite{6937196}.
\begin{figure}[htbp]
	\vspace{-0.3cm}  
	\centering
	\includegraphics[width=250pt]{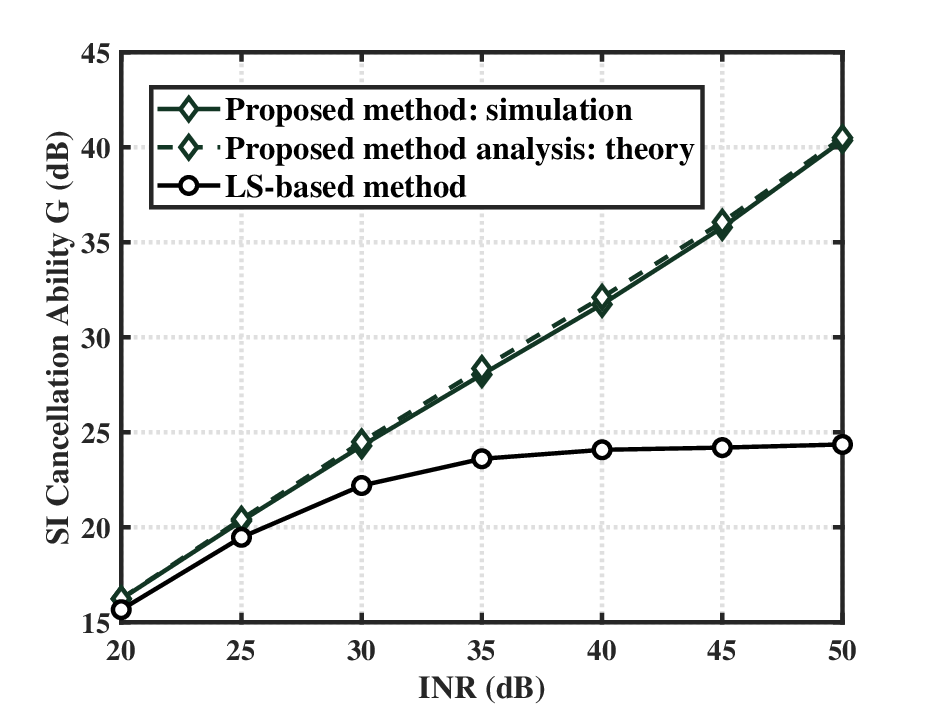}
	\setlength{\abovecaptionskip}{0cm}   
	\setlength{\belowcaptionskip}{0cm}   
	\caption{Digital SI cancellation ability versus INR, with the SNR set as 10 dB and the relative pase noise bandwidth $\Delta f=10^{-3}$.}
	\label{figure1}
	\end{figure}
\begin{figure}[htbp]
	\vspace{-0.3cm}  
	\centering
	\includegraphics[width=250pt]{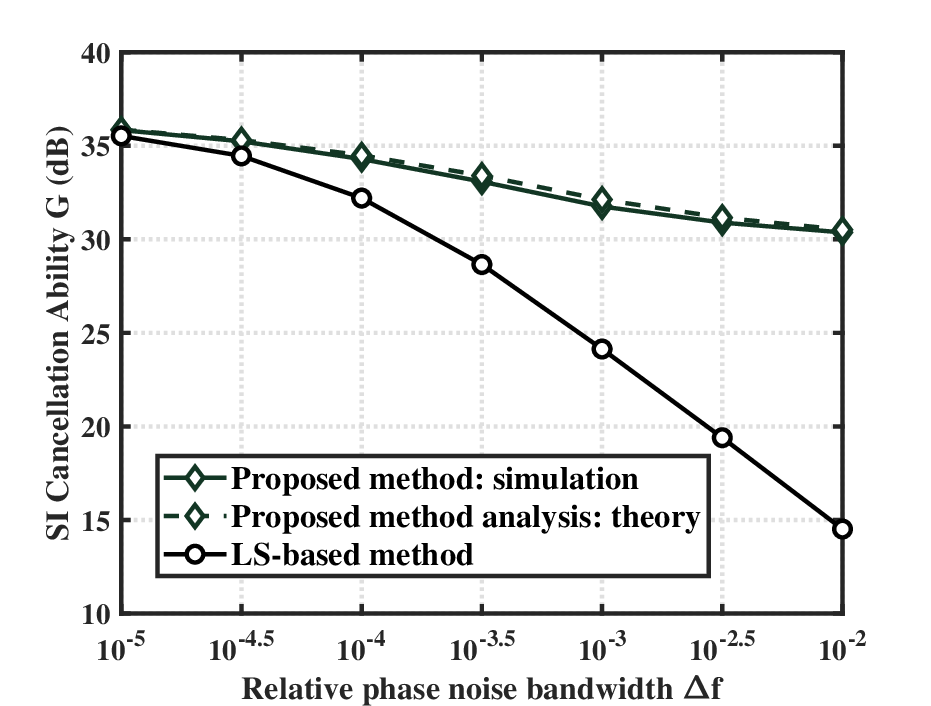}
	\setlength{\abovecaptionskip}{0cm}   
	\setlength{\belowcaptionskip}{0cm}   
	\caption{Digital SI cancellation ability versus relative phase noise bandwidth $\Delta f$, with the SNR set as 10 dB and the INR set as 40 dB.}
	\label{figure2}
\end{figure}

Fig. \ref{figure1} compares the cancellation ability of the proposed linear SI channel estimator and the conventional LS based method with different INR. As shown in this figure, the proposed method is better than the LS method in all the INR region. Besides, it is observed that with INR increasing from 20 dB to 50 dB,  the cancellation ability of the LS method increases with a gradually slow growth and finally reaches its limit as about 25 dB. This is because the LS method ignored the statistics information of the oscillator phase noise and thus its cancellation ability is limited by the ICI of phase noise \cite{7815419}. In contrast, the cancellation ability of the proposed method grows almost linerly with respect to the INR. Meanwhile, it is observed that corresponding to a linear growth, the cancellation ability of the proposed method, which increases from 15dB to 40dB, still have a 5dB loss, which implies that the ICI effects of the phase noise cannot be perfectly cancelled.

From Fig. \ref{figure2}, it can be seen that the proposed method is better than the LS based method in terms of the SI cancellation ability in the whole relative phase noise bandwidth $\Delta f$ region. Especially, the cancelaltion ability of the LS method is sensitive to $\Delta f$ where it drops by 20 dB with $\Delta f$ increasing from $10^{-5}$ to $10^{-2}$. Similar to the analysis in Fig. \ref{figure1}, this is because the LS based method is severly influenced by the ICI effects of the phase noise, and  the ICI effects increases with the increase of $\Delta f$ \cite{7815419}. In contrast, the proposed method is stable and only drops by 5 dB with the increase of $\Delta f$. In other words, the proposed method is more robust for the low oscillator quality cases.
\section{Conclusion}
In this paper, we propose a weighted linear SI channel estimator to minimize the power of the residual SI in a FD massive MIMO OFDM system with single RF chain. The closed-form expression for the power of the residual SI is formulated, and the optimal linear SI channel estimator is computed by solving the residual SI power minimization problem. Moreover, the digital SI cancellation ability of the proposed method is computed and simulated. Simulation results reveals that the proposed method significantly exceeds the conventional LS based method in large INR and low oscillator quality scenarios.
\bibliographystyle{IEEEtran}
\bibliography{huang_WCL_Ref}
\end{document}